\documentclass{article}%
\usepackage{amsmath}
\usepackage{amsfonts}
\usepackage{amssymb}
\usepackage{graphicx}%
\begin{document}

\title{Turbulent Drag Reduction of polyelectrolyte solutions: relation with the
elongational viscosity }
\author{C.~Wagner$^{1}$, Y.~Amarouchene$^{1,2}$ P.~Doyle$^{3}$, and D.~Bonn$^{1,*}$\\$^{1}${\small \ Laboratoire de Physique Statistique, UMR CNRS 8550, Ecole
Normale Sup\'{e}rieure, }\\{\small 24 rue Lhomond, 75231 Paris Cedex 05, France}\\$^{2}${\small \ Centre de Physique Mol\'{e}culaire Optique et Hertzienne,
Universit\'{e} Bordeaux 1 (UMR 5798), }\\{\small 351 cours de la liberation, 33405 Talence, France}\\$^{3}${\small \ Department of Chemical Engineering, Massachusetts Institute of
Technology, }\\{\small Cambridge, MA 02139 }\\$^{\ast}${\small \ bonn@lps.ens.fr}}
\maketitle

\begin{abstract}
{\small We report measurements of turbulent drag reduction of two different
polyelectrolyte solutions: DNA and hydrolyzed Polyacrylamide. Changing the
salt concentration in the solutions allows us to change the flexibility of the
polymer chains. For both polymers the amount of drag reduction was found to
increase with the flexibility. Rheological studies reveal that the
elongational viscosity of the solutions increases simultaneously. Hence we
conclude that the elongational viscosity is the pertinent macroscopic quantity
to describe the ability of a polymer to cause turbulent drag reduction.}

\end{abstract}

The phenomenon of turbulent drag reduction describes the diminution of the
dissipation in turbulent flow by adding tiny amounts of polymers. In this way
for instance the amount of liquid that is transported in a pipe for a given
pressure drop can be increased significantly. This implies a wide range of
industrial applications, and consequently the effect, discovered in 1949 by
Toms \cite{Toms49}, has been studied intensively over the past decades (see
e.g. Ref. \cite{Virk75a}). In spite of this effort, the underlying physical
mechanisms are ill understood and none of the existing theories match the
existing experimental data \cite{Tabor86}. The key macroscopic property of
these solutions that is known to be significantly different from that of the
solvent is its elongational viscosity. Therefore, drag reduction is usually
attributed to the elongational viscosity; however the precise connection
remains unclear.

The elongational viscosity describes the resistance of a liquid to an
elongational flow. In Newtonian liquids the elongational viscosity is just
given by the shear viscosity times a geometrical factor of three, the Trouton
ratio \cite{Trouton06}. The addition of a small amount of polymers to the
solvent does not change the shear viscosity significantly, however the
elongational viscosity can increase by several orders of magnitude. Taking an
elevated elongational viscosity into account, a qualitative explanation of
drag reduction has been given by Landahl \cite{Landahl77}: He suggests the
elongational viscosity can suppress the occurrence of streaks - regions of
flow with a high elongation rate - in the turbulent boundary layer. These
boundary layer instabilities lead to "blobs" of fluid that are ejected from
the boundary layer into the bulk of the liquid and which generate the
turbulence. An observation that provides support for the idea that the
boundary layer is important was made by Cadot et al. \cite{Cadot98}. They
showed experimentally that drag reduction occurs only in boundary layer driven
turbulence: if no boundary layers are present, indeed no drag reduction is
observed upon addition of polymers. This implies that it is the way the
turbulence is generated by the boundary layers that is altered by the addition
of polymers.

To our knowledge, however, an explicit experimental proof of the relation
between the elongational viscosity of a given polymer solution and its ability
to cause turbulent drag reduction is still missing mainly because the
determination of the elongational viscosity of drag-reducing polymer solutions
has proven difficult. It is the aim of this publication to provide this
relation. In order to do so, we study both the elongational viscosity and
turbulent drag reduction of polymers that allow us to tune their chain
flexibility. The large elongational viscosity of dilute polymer solutions is
usually attributed to the resistance to stretching of the polymer chains in
the elongational flow field. Therefore, flexible polymers should have a higher
elongational viscosity. If in addition the phenomenon of drag reduction is
indeed related to the elevated elongational viscosity, drag reduction should
also increase with increasing chain flexibility. We show here that this is
indeed the case using polyelectrolytes, for which the chain flexibility can be
tuned by the addition of salt to the solvent \cite{poly}. In that way the
chain flexibility is altered \textit{without} changing either the polymer
chemistry or the chain length (the effects of which on drag reduction are
still ill understood).

The first polyelectrolyte solution we use are aqueous solutions of double
stranded $\lambda$ DNA (48.5 bp) at the overlap concentration $c^{\ast}$ of
$40\mu g/ml$. DNA is a model monodisperse polymer and its physical properties,
for instance the relaxation time, are very well defined \cite{Perkins94}. Our
solutions contained a $10mM$ Tris and $1mM$ EDTA buffer (alternatively we also
use a $2mM$ Tris buffer) and the NaCl concentration was varied from $0$ to
$10mM$, changing the persistence length by a factor of three and the
relaxation time from about $10$ to $30ms$. The second polyelectrolyte used was
$40\mu g/ml$ hydrolyzed Polyacrylyamide (HPAA) with a molecular weight of
$5\times10^{6}$ $amu$ in water with NaCl concentrations varying from $1mM$ to
$18mM$.

First the shear rate dependent viscosity $\eta\left(  \dot{\gamma}\right)  $
of our samples was measured by use of a standard rheometer (Reologica Stress
Tech) with a cone plate geometry that assures a laminar flow (figure 1). In
the range $20s^{-1}<\dot{\gamma}<2000s^{-1}$ all HPAA samples show shear
thinning, the effect becoming less pronounced at higher salt concentrations.
This was to be expected, as stiffer polymers show in general stronger shear
thinning. The DNA solutions showed only a slight shear thinning and no
dependence on the salt concentration could be detected. At at a shear rate of
$\dot{\gamma}=2000s^{-1}$, comparable with the highest shear rates used in the
drag reduction experiments. The shear viscosities for the aqueous HPAA
solutions varies in between $1.22mPas<\eta<1.38mPas$ with the salt
concentration and for the aqueous DNA solutions the viscosity is $\eta
\sim1.08mPas$. Our rheological measurements did not reveal any measurable
normal stresses within the resolution of $N_{1}\simeq1Pa$ of the rheometer for
the aqueous $40\mu g/ml$ solutions employed here.

To investigate turbulent drag reduction we measured the drag of a turbulent
flow of the different liquids in a Couette Cell with a gap of $\delta=1mm$.
Even if the industrial implications of turbulent drag reduction are more
likely to be pipe flow, it is experimentally much more convenient to study
Couette flow; this is in fact a rather common way to characterize polymer
solutions for their ability to exhibit drag reduction \cite{Nakken01}. A
further advantage of this setup with a high surface to volume ratio is to
probe directly boundary layer effects.

Drag reduction measurements were performed on our standard rheometer. It was
driven in controlled stress mode, and drag reduction $DR$ is defined as the
normalized difference of the drag $d$ on the inner cylinder by pure water and
by the polymer solution respectively at same Reynolds number $DR=\left(
d_{H_{2}O}-d_{pol}\right)  /d_{H_{2}O}\times100\%$. The Reynolds number is
defined as $Re=\frac{\Omega r\delta\rho}{\eta}$, $\Omega$ being the rotation
speed of the inner cylinder, $\rho$ the liquid density and $\eta$ the solvent viscosity.

Figure 2 shows the measured drag reduction as a function of time for a
Reynolds number $Re\approx2000$. Both for the DNA and for the HPAA solution
$DR$ is found to depend on the salt concentration in the solvent: drag
reduction increases with increasing salt. For the DNA\ solution the results
are qualitatively as for the HPAA, but the effects are less pronounced.

We believe the interesting temporal behavior for times $t\sim\leq400$ sec to
be a rather complicated mixture of disentanglement of different polymer chains
\cite{Cadot98}, rupture of the individual polymer chains by the strong flow
\cite{Virk75a}, and relaxation of the flow towards a steady state. In addition
to the data shown here, at longer times we observe a tendency to smaller
values of drag reduction, probably due to further degradation of the polymers
\cite{Nakken01}.

Drag reduction as a function of the Reynolds number is shown in figure 3. For
each new Reynolds number the system was given $60$ seconds to reach a steady
state, a reasonable compromise in view of the different temporal effects
discussed above. For all the solutions, we observe an increase of Drag
reduction with increasing ionic strength. We checked that this is not due to
our use of the Couette cell, or the very moderate Reynolds numbers employed
here by repeating the experiments in the turbulence cell of Cadot et al.
\cite{Cadot98}. These high-Reynolds number experiments gave very similar
results. The highest Drag reduction measured with HPAA for a $Re\approx
4.10^{5}$ was $DR\sim30\%$ rather close to the drag reduction $DR\sim20\%$
measured with the Couette cell. The complete study of drag reduction at high
Reynolds number will be published elsewhere \cite{Yacine}.

Having measured the effect of salt on drag reduction\footnote{In a recent
paper, Choi et al. \cite{Choi} have reported on the drag reduction properties
of DNA at lower concentrations but higher Reynolds numbers . However, their
focus was on studying the degradation introduced by the high stretching fields
and no correlation with the elongational viscosity was given.}, we now turn to
the elongational viscosity. As the velocity fluctuations in turbulence are
violent, and rapidly change direction, it is unlikely that the polymer
extension in such flows reaches a steady state. The pertinent experiment is
therefore to study incipient (startup) elongational flow, and measure the
response of the polymer solutions to that .

In order to measure the dynamic elongational viscosity $\eta_{e}(t)$ of our
solutions we followed the method proposed by Bazilevskii et al.
\cite{Bazilevskii81} and further developed by Amarouchene et al.
\cite{Amarouchene01}. Therein the authors describe the detachment process of a
droplet of a diluted polymer solution from a capillary. The addition of a
small amounts of flexible polymers to water inhibits the finite time
singularity break up \cite{Papageorgiou95} of the droplet, and a cylindrical
filament is formed between the capillary and the droplet. The flow profile in
the filament is purely elongational and by balancing the capillary forces to
the elastic stresses, the elongational viscosity $\eta_{e}(t)$ can be
extracted from the measured filament diameter $h(t)$. The elongational
viscosity $\eta_{e}(t)$ is time dependent because in the course of the
experiment the polymers are more and more stretched by the flow. The
experiments reveal that the temporal behavior of $h(t)$ is exponential,
implying a unique thinning regime for which the elongational rate
$\dot{\epsilon}=-2\frac{\partial_{t}h}{h}$ remains constant.

Recently Anna and Mc Kinley \cite{Anna00} have shown that filament thinning
techniques analogous to the one used here lead to measures of the elongational
viscosity that are consistent \ with the more established filament stretching
techniques. The latter cannot be employed here since it only allows for
measurements for much more viscous solvents (typically $1Pa.s$) \cite{Anna00}.

It turns out that for the aqueous solutions used in the $DR$ experiments, the
droplet detachment process is too fast to observe, even using a rapid camera
(Photonetics) working at a 1000 frames/s. Increasing the solvent viscosity
slows down the dynamics of the filaments sufficiently so that these can be
followed, and the elongational viscosity can be determined. We will show below
that these measurements suffice also to deduce the elongational viscosity of
the aqueous solutions used in the drag reduction experiments.

The solvent viscosity could be adjusted by adding a certain amount of
glycerol. Figure 4 shows the photographs of the detachment process for the DNA
solution for the $0mM$ and the $10mM$ salt concentration. The trend can be
observed directly on the pictures: the more flexible polymer has a higher
elongational viscosity; it is evident from the figure that the filament
thinning is much slower.

In the filaments, for the DNA solutions, the elongational rate varies with
$NaCl$ : $\dot{\epsilon}=90s^{-1}$ for low concentration and $\dot{\epsilon
}=50s^{-1}$ for the high concentration. In order to be able \ to compare
different experiments,\ it is thus useful to introduce the so called Hencky
strain, the product of the elongation rate and the elapsed time $\dot
{\epsilon}t$, which measures the total deformation of the fluid element
containing the polymer. We define the zero of the Hencky strain for the moment
when the filament is formed, i.e. when the flow is strong enough to interact
with the polymers \cite{Amarouchene01}.

Figure 5 shows the elongational viscosities of the different samples, for
different salinities as a function of the Hencky strain. Again both for the
DNA and the HPAA, $\eta_{e}$ depends strongly on the salt. For HPAA, $\eta
_{e}$ changes more than a factor of three between the lowest and the highest
salinity, while there is still a factor of almost two observed for the DNA solution.

A prediction of all polymer flow models is that the elongational viscosity
scales linearly with the solvent viscosity \cite{bird}. If this is true, we
can directly extract the elongational viscosity of the aqueous solutions used
in the drag reduction experiments from the elongational viscosities of the
glycerol-water solutions. To verify this, we prepared a series of additional
samples of $300\mu g/ml$ PAA in different glycerol-water mixtures. This allows
us to change the solvent viscosity $\eta$ in the range $10mPas\leq80mPas$. The
inset of figure 5 shows the Trouton ratio $Tr$, the measured elongational
viscosity $\eta_{e}$ of the polymer solutions divided by three times the
viscosity of the solvent ($Tr/3\equiv1$ for Newtonian liquids). The Trouton
ratio of the different solutions perfectly collapses on to a single master
curve. For even smaller viscosities slight discrepancies can be observed, but
here the dynamics becomes again too fast to allow for a precise determination
of $\eta_{e}$. The perfect collapse also indicates that the chain flexibility
is not modified by the addition of different amounts of glycerol. Therefore we
can obtain the elongational viscosities $\eta_{e}$ of the aqueous polymer
solutions by dividing the data from the glycerol water solutions by the ratio
of the solvent shear viscosities.

It is therefore now possible to relate the drag reduction directly to the
elongational viscosity. The result is shown in figure 6. We plot the drag
reduction for a given Reynolds number as a function of the elongational
viscosity $\eta_{e}$ at a Hencky strain of $1$. Drag reduction is shown to
increase monotonically with the elongational viscosity for a given polymer and
the data for the different polymers collapse, demonstrating that indeed the
elongational viscosity is the pertinent macroscopic quantity to account for
$DR$. As far as we can tell, this is the first explicit demonstration that the
turbulent drag reduction is directly connected with the elongational viscosity
of a polymer solution.

These results, especially those on DNA, open the way to a microscopic
understanding of both the enormous elongational viscosity dilute polymer
solutions can have, and the surprising results thereof: turbulent drag
reduction. It has recently become possible to observe the deformation of the
polymer chains due to flow of single DNA molecules using fluorescence
microscopy \cite{Perkins94}. Performing such an experiment on the filaments
would allow to relate the extension of the individual polymer chains, to the
macroscopic stresses the chain extensions generate. If subsequently the
polymer conformation is visualized in a turbulent boundary layer, an estimate
of the extra stresses in the boundary layer could be obtained, which could be
the key to the understanding of this phenomenon. These experiments are in progress.


\textbf{Acknowledgements}\textit{\/} LPS de l'ENS is UMR 8550 of the CNRS,
associated with the universities Paris 6 and 7. We thank J. Meunier and J.
Eggers for discussions. This work is supported by the EEC IHP-MCFI program.



\textbf{figure 1}: The shear dependent viscosity of the different polymer
solutions. Filled symbols: aqueous HPAA solutions $1mM$ NaCl (circles),
$2mM$(up triangles), $7mM$ (down triangles), $18mM$(diamonds). Filled symbols:
aqueous DNA solution $0mM$ NaCl (up triangles). \newline

\textbf{figure 2}: The turbulent drag reduction $DR$ as function of time at a
constant Stress of 8 Pa yielding a Reynolds Number $Re$ $\approx2000$. Filled
symbols: DNA solutions $10mM$ NaCl (down triangles), the other symbols like in
figure 1. The error bars indicate the statistics of three consecutive
measurements. \newline\newline\textbf{figure 3}: The turbulent drag reduction
$DR$ as a function of the Reynolds number $Re$ for $40\mu g/ml$ HPAA a) and
$40\mu g/ml$ \ DNA solution.\newline

\textbf{figure 4}: The detachment process of the droplet for the $40\mu g/ml$
DNA solutions with $80/20vol\%$ glycerol-water as solvent and a),b),c),d)
$0mM$ Nacl and e),f),g),h) $10mM$ NaCl. The rows show different time steps $t$
in regard to the moment $t_{c}$ when the filament breaks. a),e): $t_{c}%
-120ms$; b),f): $t_{c}-94ms$; c),g): $t_{c}-53ms$; d),h): $t_{c}-20ms$. The
viscosity of the solvent is $80$ $mPas$ and the diameter of the capillary used
to study the droplet formation has a diameter of $1.2mm$. \newline

\textbf{figure 5}: The elongational viscosity $\eta_{e}$ for the different
polymer solutions. The symbols are like in figure 1, only the solvent is a
$80/20vol\%$ glycerol-water mixture. Inset: The Trouton ratio $Tr$ for
solutions with $300\mu g/ml$ PAA in different water glycerol mixtures and
$0mM$ salt. The solvent viscosities are $\eta$= $80$ (circles), $43$
(squares), $20$ (up triangles), $10$ (diamonds) $mPas$. \newline

\textbf{figure 6}: The drag reduction DR at a Reynolds number $Re_{poly}=1400$
as a function of the elongational viscosities of the aqueous polymer
solutions, for different salinities at a Hencky strain of $\dot{\epsilon}t=1$.
To allow for a comparison between samples with a different shear viscosities,
the Reynolds number $Re_{poly}$ is calculated using the laminar shear
viscosities of the polymer solutions at a shear rate $\dot{\gamma}=2000$.
Filled squares: DNA-solutions, open squares: HPAA solutions.\newline


\begin{thebibliography}{99}                                                                                               %


\bibitem {Toms49}B.~A.~Toms, Proc. Int. Cong. Rheol. \textbf{II}, (1949) 135.

\bibitem {Virk75a}J.~L.~Lumley, Ann. Rev. Fluid Mech. \textbf{1} (1969) 367;
P.~S.~Virk, AIChE \textbf{21}, 625 (1975); A.~Gyr and H.~W.~Bewersdorff, Drag
reduction of turbulent flow by additives (Kluwer, Dordrecht, 1995);
R.~Govindarajan, V.~S.~L%
\'{}%
vov, I. Procaccia, Phys. Rev. Lett. \textbf{87} (2001)174501; The dependence
of DR on the salt concentration of a polyelectrolyte solutions was also
already reported by Virk in 1975 : P.~S.~Virk, Nature \textbf{253} (1975) 109.

\bibitem {Tabor86}M.~Tabor and P.~G.~de~Gennes, Europhys. Lett \textbf{2}
(1986) 519; J.~M.~J.~Den~Toonder, M.~A.~Hulsen, C.~D.~C.~Kuiken,
F.~T.~M.~Nieuwstadt, J. Fluid. Mech. \textbf{337 }(1997) 193; R.~Sreenivasan,
C.~M.~White, J. Fluid Mech. \textbf{409} (2000) 149.

\bibitem {Trouton06}F.~T.~Trouton, Phil. Mag. \textbf{19} (1906) 347.

\bibitem {Landahl77}M.~T.~Landahl, Phys. of Fluids \textbf{20} (1977) 55.

\bibitem {Cadot98}O.~Cadot, D.~Bonn, S.~Douady, Phys. Fluids \textbf{10
}(1998) 426 .

\bibitem {poly}C.G. Bauman, S.B. Smith, V.A. Bloomfield, C. Bustamante,
P.N.A.S. \textbf{94} (1997) 6185.

\bibitem {Perkins94}T.~T.~Perkins, S.~R.~Quake, D.~E.~Smith, S.~Chu, Science
\textbf{264} (1994) 823.

\bibitem {Nakken01}V.~N.~Kalashnikov, J. Non-Newt. Fluid Mech. \textbf{75}
(1998) 209. T.~Nakken, M.~Tande, A.~Elgsaeter, J. Non-Newt. Fluid Mech.
\textbf{97 }(2001) 1.

\bibitem {Yacine}Y.~Amarouchene, C. Wagner, D.~Bonn, to be published.

\bibitem {Choi}H.~J.~Choi, S.~T.~Lim, P.~Lai, C.~K.~Chan, Phys. Rev. Lett.
\textbf{89} (2002) 088302.

\bibitem {Bazilevskii81}A.~V.~Bazilevskii, S.~I.~Voronkov, V.~M.~Entov,
A.~N.~Rozhkov, Phys. Dokl. \textbf{26} (1981) 333.

\bibitem {Amarouchene01}Y.~Amarouchene, D.~Bonn, J.~Meunier, H.~Kellay, Phys.
Rev. Lett. \textbf{86} (2001) 3558.

\bibitem {Papageorgiou95}J.~Eggers, Rev. Mod. Phys. \textbf{69} (1997) 865.

\bibitem {Anna00}S.~L.~Anna, G.~H.~McKinley, J. Rheol. \textbf{45} (2001) 115.

\bibitem {bird}R.~B.~Bird, R.~C.~Armstrong, O.~Hassager, Dynamics of Polymeric
Liquids (Wiley, NY, 1987) Vols. I and II.
\end{thebibliography}
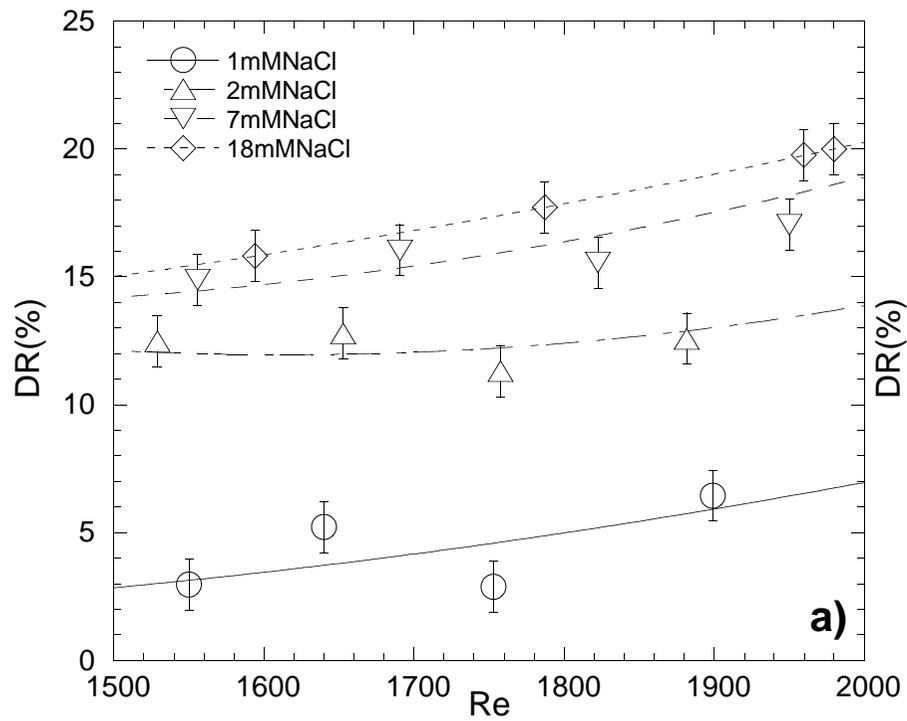
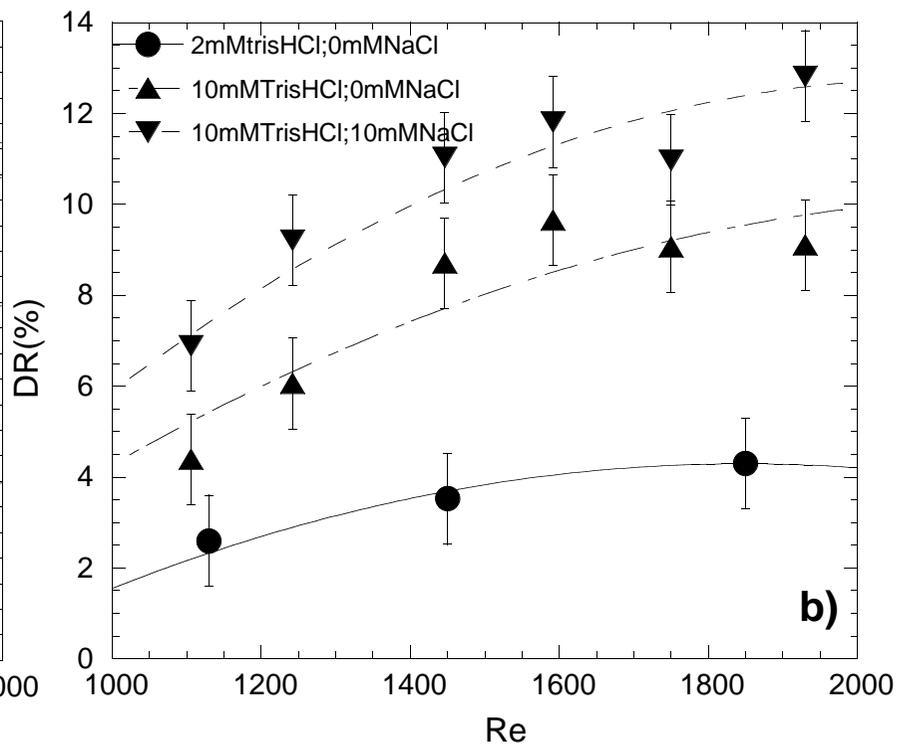
\end{document}